# APPLYING PART-OF-SPEECH ENHANCED LSA TO AUTOMATIC ESSAY GRADING

Tuomo Kakkonen[1], Niko Myller, Erkki Sutinen

Department of Computer Science
University of Joensuu
Joensuu, Finland
firstname.lastname@cs.joensuu.fi

*Abstract* Latent Semantic Analysis (LSA) is a widely used Information Retrieval method based on "bag-of-words" assumption. However, according to general conception, syntax plays a role in representing meaning of sentences. Thus, enhancing LSA with part-of-speech (POS) information to capture the context of word occurrences appears to be theoretically feasible extension. The approach is tested empirically on a automatic essay grading system using LSA for document similarity comparisons. A comparison on several POS-enhanced LSA models is reported. Our findings show that the addition of contextual information in the form of POS tags can raise the accuracy of the LSA-based scoring models up to 10.77 per cent.

## I. INTRODUCTION

Latent Semantic Analysis (LSA) is a method for document similarity comparison [1]. It has been successfully applied to several *Information Retrieval* (IR) tasks, such as information filtering [2] and document classification [3]. A successful application area for LSA has been educational technology, where it has been used for assessing summaries [4], providing automated feedback in intelligent tutoring systems [5,6], and as in this paper, for automatically grading student essay responses [7,8].

The LSA model is based on "bag-of-words" assumption, thus it does not take the internal structure of sentences or word ordering into consideration. When LSA is applied, a *word-by-context matrix* (WCM) representing the number of occurrences of each distinct word in each context (*i.e.* sentence, paragraph or document) is constructed with lemmatized words.

While the exact link between syntax and semantics in natural language is a topic of dispute, the fact that there is a connection between these two levels is hardly controversial. Although LSA has proven to be a suitable model for comparing document similarities based on word co-occurrences in global contexts, such as the structures spanning over several sentences, paragraphs or documents, it lacks the sensitivity for a local context, such as the internal structure of sentences. Thus, a model integrating local as well as distant relations between words appears as an interesting method for document modeling.

Our aim in this paper is to explore the ways in which *part-of-speech* (POS) information can be used to enhance the LSA model to add local information about the internal relations between the words in sentences. The most straight-forward use of POS information is to use the POS tag of each word in addition to the base form of the word when building the WCM. Adding POS tags to LSA model disambiguates the meaning between words with the same base form but different POS tag.

The basic LSA model lacks sensitivity to the context in which the words occur. We tackle the problem by adding the POS tags of the preceding and succeeding words to form a local context for each word occurrence.

In addition, POS tags can be used for filtering out some word classes from the matrix. For example, one might want to use only the *content words* (*i.e.* nouns, verbs and adjectives) in the matrix in order to reduce the size of the resulting WCM, thus reducing the time and space complexities of the model computation. We also experiment with an LSA model that adds to the WCM all the word forms left ambiguous by the POS tagger and parser. Finally, we introduce two models that combine models just introduced. In these models, two or three WCM entries are added for each word occurrence.

We report a comparison of several POS-enhanced LSA models. The accuracy of these models is examined with an automatic essay grading system. The approaches, in which more than one entry for a word occurrence is added to the WCM has not, to our knowledge, been previously introduced.

Section 2 describes previous work in LSA-based automatic grading and methods to enhance the LSA model with syntactic information. Section 3 explains LSA briefly and the essay grading system, *Automatic Essay Assessor* (AEA), is introduced. Section 4 describes the test sets, the configurations of the experiments and summarizes the results. Section 5 concludes the findings, with directions for future research.

[1] The research reported in this paper has been supported by the Automated Assessment Technologies for Free Text and Programming Assignments project funded by the Academy of Finland. Kakkonen was working at the Faculty of Philosophy, University of Split, Croatia while working on this paper.

*A. Definitions*

We refer to a written composition answering an examination question as an *essay* or essay document. An *essay prompt* is a description of the assignment for the writers. A *word* occurs in a *context*, for example in a sentence, a paragraph, or a document. The words in each context can be represented as a *document/query vector* containing the number of occurrences of each word in the context for every distinct word in the *corpus*. These vectors can be combined to form a WCM that is an $m*n$ matrix where $n$ is the number of documents and $m$ is the number of distinct words in the corpus. The *base form* or lemma of the word is simplified form (the part of the word in which the affixes are attached) of the word returned by the parser or stemmer. *Part-of-speech tag* (POS tag) is the representation of the POS of the word returned by the parser. For convenience, we use *N* for nouns, *V* for verbs and *A* for adjectives.

## II. Previous Work

Several automatic essay grading systems apply LSA. For example Landauer et al. [9] and Folz et al. [10] have reported correlations of 0.64…0.84 between grades given by two human assessors and correlations of 0.59…0.89 between the LSA-based grading system and human graders. Thus, it can be argued that LSA-based grading systems outperform or at least perform as well as the human graders.

Several approaches have been applied in order to enhance LSA with syntactic and morphological knowledge. Wiemer-Hastings and Zipitria [11] add the POS tag of each word into the WCM, calling the model *Tagged LSA*. The aim of the model is to distinguish between words that are used in multiple parts-of-speech. Tagged LSA was applied as a part of an intelligent tutoring system. In Hastings and Zipitria's experiments the Tagged LSA model lowered the correlation between the human and computer feedback. They suspected a reason being that the stochastic tagger they applied was not trained with suitable data, thus making errors in assigning POS tags. Wiemer-Hastings [12] introduced another model, in which the sentences were separated into atomic clauses and propositions before feeding them into the LSA model. Sentences were manually segmented into subject and object phrases and verb phrases. In addition, anaphora was resolved by replacing pronouns by their antecedents. The experiments showed that such method performed worse than the basic LSA model. In another experiment, Wiemer-Hastings and Zipitria [11] enhanced the LSA model by adding syntactic roles, such as subject, predicate and object, and showed that the approach, called *Structured LSA* (SLSA) outperformed basic LSA.

Closest to the work represented in this paper is the *Syntactically Enhanced LSA* (SELSA) model by Kanejiya et al. [13,6] that has been applied to both language modeling and as a part of an intelligent tutoring system. In SELSA, the POS tag of the preceding word is used to model the syntactic context of a word. For each word occurring in the text, the word concatenated with the POS tag of the previous word is added to the WCM. The findings in intelligent tutoring domain were ambiguous. The SELSA model was able to correctly evaluate a few more student answers than LSA. On the other hand, the correlation between the system and human-given feedback got lower. In language modeling, using LSA resulted in a model with somewhat lower perplexity than the one created with SELSA. Thus, SELSA was slightly outperformed by LSA.

## III. Brief Introduction to LSA and the Essay Grading System

*A. LSA*

LSA is a method for determining the similarity of the meaning of words and text passages based on word co-occurrence data. The basic assumption is that a close relationship holds between the meaning of a text and the words in that text. LSA is often able to detect the similarity of two texts even though the documents do not contain common words, thus providing a way to account for synonymy and polysemy.

When LSA model building is started, the text is first represented as a WCM. Next, standard IR preprocessing steps are applied, most commonly consisting of lemmatization or stemming of words, entropy-based term weighting, and removal of stopwords and words occurring only once.

The essence of LSA lays in the dimensionality reduction step. In dimensionality reduction, the dimension of the WCM is lowered by applying *singular value decomposition* (SVD) and then reducing the number of singular values in the SVD. The process increases the dependency between contexts and words, making the underlying semantic structure to become evident by reducing the noise in the dyadic data. More detailed descriptions of LSA may be found in *e.g.* [1].

*B. Automatic Essay Assessor*

*Automatic Essay Assessor* (AEA) [8] is an automatic essay grading system. The system consists of three main components: A syntactic *parser* and morphological analyzer which is currently based on the *Constraint Grammar* (CG) in order to find the base form and POS tag for each word [14, 15]. LSA or some other *IR model* is applied in order to measure the similarity between the essays and the course materials. The *scoring model* of AEA is based on both the course content (textbook passages, lecture notes *etc.*) and human-graded essays.

The assessment procedure consists of two phases. First, the course materials representing the essay prompt specific knowledge is processed to form the WCM. The WCM is then given as input to either LSA or some other IR model. Next, AEA uses human-graded essays to determine the threshold similarity values for each grade category by comparing essays to the LSA model created from the course materials. A query vector representing the content of the essay is created and compared to each document of the LSA model and similarity

**Table 1. The test sets used in the experiments.**

| Set No. | Domain | Level | Train. essay | Test essays | Grade scale | Grader | No. sent. | No. words |
|---|---|---|---|---|---|---|---|---|
| 1 | Education | Undergrad. | 70 | 73 | 0-6 | Professor | 147 | 2397 |
| 2 | Communications | Vocational | 42 | 45 | 0-4 | Course teacher | 139 | 1583 |
| 3 | Soft. Eng. | Graduate | 26 | 27 | 0-10 | Assistant | 105 | 965 |

values are summed to get a similarity score for the essay.

In the grading phase, a document vector is created for each essay to be graded. To grade an essay, its document vector is compared to the reference materials with the same method that was applied in the previous phase. The similarity score of the essay is then matched to the grade categories according to their limits to determine the correct grade. More detailed descriptions of the system can be found from [8,16].

So far, we have applied the system only to essays written in Finnish, although the system is not limited to only one language. In order to assess essays in another language, it is only required to change the parser and stopword list to conform to the used language.

*C. Part-of-speech Enhanced LSA (POSELSA)*

In our *Part-of-speech Enhanced LSA* (POSELSA) model the WCM is composed in three ways by adding to the model the base form of the word together with:
– the POS tag of the current word (POS);
– the POS tag of the current and preceding word (Prev. POS); and
– the POS tag of the current and succeeding word (Next POS).

For example, for the Finnish phrase "puolustuksen suomalainen tukipilari" ("the Finnish corner-stone of the defense", *lit*. "defense's Finnish pillar") the following three entries would be added to the WCM in a basic LSA model: "puolustus", "suomalainen", "tukipilari". In the model including the POS tag of the current word, the tags N (noun), A (adjective) or N would be combined with each of the three words, respectively. In the model using the POS tags of the current and preceding words, the following base form and tag pairs would be used: (PUNCT, "puolustus", N), (N, "suomalainen", A), and (A, "tukipilari", N). For instance, the first POS tag and base form group denotes that the word, whose base form is "puolustus", is a noun (N) and is preceded by a punctuation (PUNCT), thus starting a new sentence or a phrase separated from the previous one by a comma or period.

In addition, we experiment with the model by adding all the word forms that are left disambiguated by the FINCG tagger/parser (*Amb.*). For example, let the parser return for the word form "häntä" two possible base form-POS tag combinations: "häntä, N" (tail) and "hän, P" (him/her, pronoun, partitive case). In the classical LSA model only the first base form "häntä" would be added into the model. In the model that preserves the ambiguous forms, both lemmas along with their POS tags are added into the WCM.

IV. EXPERIMENT AND RESULTS

The accuracy of the POSELSA models were compared against the basic LSA model with the three tests sets introduced in Table 1. The domain column indicates the subject of the essays and the column level tells if the essays were collected from an undergraduate, a graduate course or a vocational school examinations. The next two columns show the number of essays used for creating the scoring model and the number of essays graded by the system. Grade scale indicates the scope of grades used by the system and the grader, who is also stated in the next column. The last two columns indicate the total number of sentences and words in the course material corpus.

With each POSELSA model, the grading was performed with all the possible dimensions ranging from 2 to the number of contexts in the WCM. In Table 2, the Spearman correlations between the grades given by the system and the human assessor are reported for the dimension that resulted in the most accurate grading. For each of the models, the average Spearman correlation over the three test sets, weighted based on the number of graded essays in each set, is reported with the percentage difference to the baseline model. In the baseline model (denoted by #) the WCM is constructed using only the lemmas of each word. In the two first models reported in the first row, just the lemma (the baseline, column *Lemma*) or lemma with the POS tag of the current word (column *POS*) is added to the WCM. In the models reported in the next two columns, in addition to the lemma and POS tag of the current word, also the POS tag of the previous word (*Prev. POS*) or the next word (*Next POS*) is added to the WCM. The second row (*Amb.*) reports the results for the models in which all the word forms left ambiguous by the parser are added to the model. The third and fourth rows show the results of the models that use only the content words (N, V, A) for building the WCM.

Results indicate that the highest accuracy was achieved by adding the POS tag of the current and succeeding word to the lemma with all ambiguities added as well. However, almost all of the reported models resulted in higher accuracy compared to the baseline. We also run experiments to test if using several entries for a single word instance further improves the results. In this way, the syntactic context of the word occurrence can be better defined.

**Table 2. The results of the comparison between the POS-enhanced LSA models. The accuracy is reported as the average Spearman correlation over the three essay sets. Result of the baseline model is marked with #.**

| Model | Lemma | Diff. % | POS | Diff. % | Prev. POS | Diff. % | Next POS | Diff. % |
|---|---|---|---|---|---|---|---|---|
| **All** | 0.7652 # | - | 0.7928 | 3.60 | 0.72003 | -5.86 | 0.8121 | 6.13 |
| **All + Amb.** | 0.7789 | 1.80 | 0.7817 | 2.16 | 0.7607 | -4.16 | 0.7810 | 2.07 |
| **Cont. words** | 0.7947 | 3.86 | 0.8171 | 6.78 | [2] | -5.90 | [2] | -2.84 |
| **Cont. words + Amb.** | 0.8016 | 4.76 | 0.8007 | 4.65 | [2] | -11.95 | 0.8476 | 10.77 |

The models introduce a separate entry for each base form / POS tag combination. In the first model (C+N), each word occurrence in a text is modeled with two entries in the WCM, consisting of the base form of the word together with:
– the POS tag of the current word (POS) (as in the POS model in the previous experiment);
– the POS tag of the current and preceding word (as in the Prev. POS).

In the P+C+N model a third entry is added, which combines the base form and POS tags of the current and succeeding words. Table 3 reports the accuracy of the models.

**Table 3. The results of the comparison between the LSA models. The format of the table and the baseline measure are the same as in Table 2.**

| Model | C+N | Diff. % | P+C+N | Diff. % |
|---|---|---|---|---|
| **All POS** | 0.8015 | 4.75 | 0.8100 | 5.85 |
| **All POS + Amb.** | 0.8101 | 5.87 | 0.8114 | 6.04 |
| **Cont. words** | 0.8308 | 8.58 | 0.8087 | 5.68 |
| **Cont. words + Amb.** | 0.8147 | 6.47 | 0.8290 | 8.35 |

Both the two models resulted in improved grading accuracy compared to the baseline model, the latter model performing better. Moreover, as with the models that use a single entry per word occurrence, including ambiguous words resulted in higher accuracy.

---

[2] The correlation is not based on all the three test sets. Due to the removal of word occurrences with only a single instance in the WCM, the models of one or two test sets resulted in a WCM with more contexts than words and SVD could not be computed. Thus, the correlation is not comparable to the other test runs. In those cases, only the difference to the baseline model is reported, based on the test sets that could be used.

On the other hand, building these models results into extremely large WCMs. In the P+C+N model, each word has at least three values in the WCM. With the ambiguous words included, the number of entries per word occurrence is even higher. Thus, the applicability of these models is limited to relatively small document collections (up to several thousand documents). SEQARABIC

## V. CONCLUSION

We have reported experiments in using LSA models that incorporate POS tags into the semantic space. We have applied to models in the context of automatic essay grading. Using POS tags to filter out all other words except the content words improves the accuracy and at the same time decreases the time and space it takes to compute the LSA. Furthermore, the results indicate that syntactic information in the form of POS tags increases the accuracy of the system by about five to ten per cent.

The best model uses only the content words concatenated with the POS tag of the succeeding word, and all ambiguous forms of each word. In general, most of the models using the syntactic information about the context of the word, *i.e.* the POS tag of the previous and next word, resulted in higher accuracy compared to the baseline measure.

Combining several of these models into single model that represents each word occurrence with several WCM entries improved the accuracy from 4.75 to 8.58 per cent compared to the baseline model. However, the computational complexity of these models may hamper their applicability.

Compared to the previous experiments in enhancing LSA with morphological and syntactic information, our results are better. However, as the other models have not been applied to automatic essay grading, direct comparisons could not be performed. Furthermore, no manual preprocessing is needed in our model, as in the case of Tagged LSA.

As a future research, we expand the results into English in the context of Information Retrieval. In addition to giving more insight to the generalizability of the models, such experiments on larger document collections would allow us to perform statistical significance tests.

We plan to incorporate different kinds of syntactic information to the model in the form of subject, predicate and object relations. Another direction for further research is to incorporate syntactic information into probabilistic language models such as Probabilistic LSA and Latent Dirichlet Allocation [16].

Finally, the POS-enhanced models offer an interesting domain for comparative tagger and parser evaluation. Comparing the results of several tagging and parsing systems in the POS-enhanced LSA models would give insight on their practical accuracy.